\documentclass[%
 reprint,
 amsmath,amssymb,
 aps,
]{revtex4-2}

\usepackage{bbm}

\usepackage{lineno}

\usepackage{relsize}

\usepackage{graphicx}
\usepackage{dcolumn}
\usepackage{bm}

\usepackage{xcolor}
\newcommand{\ii}{\mathrm{i}}

\begin{document}

\preprint{APS/123-QED}

\title{Analog dual classifier via a time-modulated neuromorphic metasurface}

\author{M. Mousa$^{1}$}
\author{M. Moghaddaszadeh$^{1, 2}$}
\author{M. Nouh$^{1,2}$}\altaffiliation[Corresponding author: ]{mnouh@buffalo.edu}

\affiliation{\vspace{2ex} \makebox[\linewidth][c]{{\normalfont $^1$}Dept. of Mechanical and Aerospace Engineering, University at Buffalo (SUNY), Buffalo, NY 14260-4400}\\
\makebox[\linewidth][c]{{\normalfont $^2$}Dept. of Civil, Structural and Environmental Engineering, University at Buffalo (SUNY), Buffalo, NY 14260-4300}\vspace{2ex}}

\date{\today}

\begin{abstract}
A neuromorphic metasurface embodies mechanical intelligence by realizing physical neural architectures. It exploits guided wave scattering to conduct computations in an analog manner. Through multiple tuned waveguides, the neuromorphic system recognizes the features of an input signal and self-identifies its classification label. The computational input is introduced to the system through mechanical excitations at one edge, generating elastic waves that traverse multiple layers of resonant metasurfaces. These metasurfaces possess a tunable phase akin to trainable parameters in deep learning algorithms. While early efforts have been promising, the well-established constraints on wave propagation in finite media limit such systems to single-task realizations. In this work, we devise a dual classifier neuromorphic metasurface and demonstrate its effectiveness in carrying out two completely independent classification problems that are concurrently carried out in parallel, thus addressing a major bottleneck in physical computing systems. Parallelization is achieved through smart multiplexing of the carrier computational frequency, enabled by prescribed temporal modulations of the embedded waveguides. The presented theory and results pave the way for new paradigms in wave-based computing systems, which have been elusive thus far.
\end{abstract}

\maketitle


\section{Introduction}
\label{introduction}

\begin{figure*}[t!]
\centering
\includegraphics[width=\linewidth]{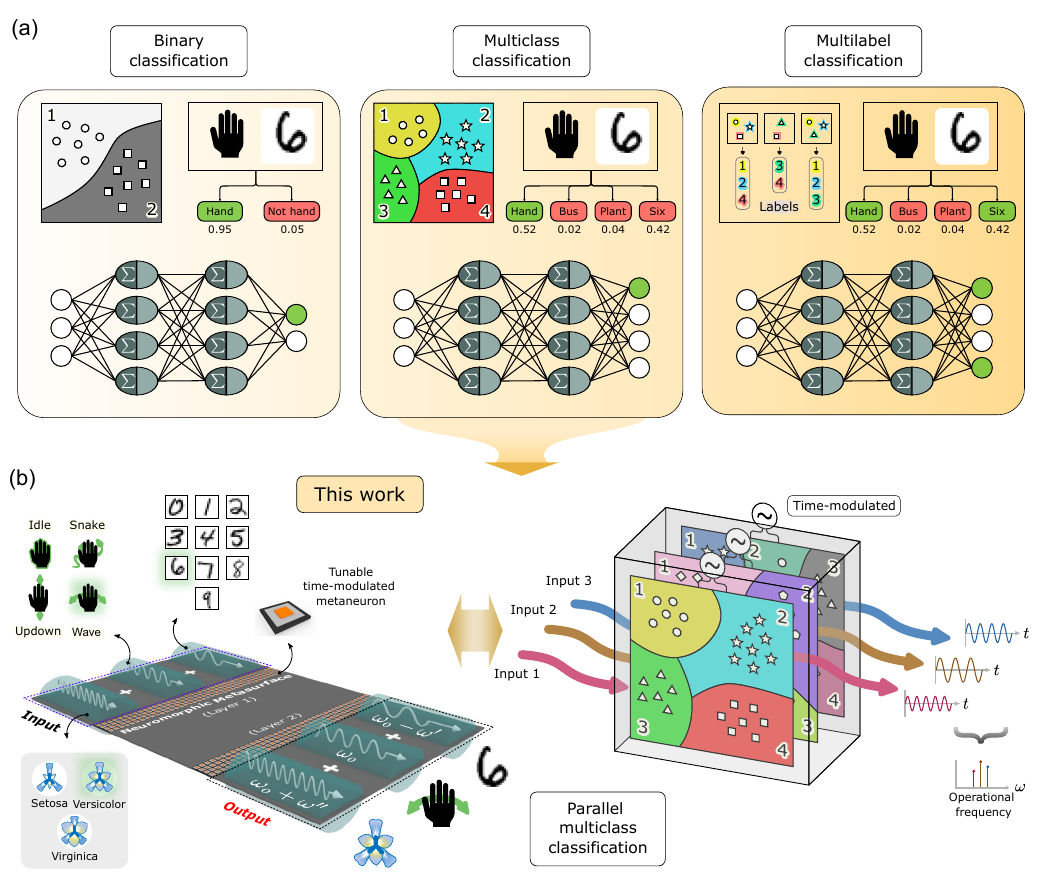}
\caption{(a) Popular classification problems. Binary classifiers distinguish between two classes: multiclass classifiers categorize inputs into one of several classes; and multilabel classifiers, which can assign multiple labels to a single input. (b) A schematic of the proposed dual classifier neuromorphic metasurface. The left panel shows a system consisting of two time-modulated metasurfaces, capable of independently classifying samples from three distinct datasets. The right panel illustrates the system's ability to function as three separate multiclass classifiers, while being structurally a single neuromorphic metasurface system.}
\label{Fig1}
\end{figure*}

Driven by rapid advancements in metamaterial design and fabrication techniques, mechanical computing has garnered significant attention over the past decade \cite{yasuda2021mechanical}. Contrary to digital systems where a signal varies virtually via discrete values of both time and amplitude, analog computing (of which mechanical computing is a subset) relies on the continuous variation of a physical phenomenon to correlate an output of a computational task to an input that takes the same physical form \cite{maclennan2007review}. This allows readily available information, in the form of external vibroacoustic loads or temperature changes, or even internally through backscattering from structural defects or local inhomogeneities, to directly stimulate the system, effectively giving it a degree of acclimatization with the surrounding environment. Perhaps more importantly, the ability to infuse computational and decision-making capabilities into a self-contained structure with minimal energy requirements renders such systems highly resilient in extreme environments in which digital processing of information is not feasible.  

Motivated by this promising vision, there has been a spurt of research activity seeking various mechanical systems that are capable of realizing basic computations. Among these are systems that carry out logic and Boolean operations via origami \cite{treml2018origami,liu2023discriminative}, conductive metamaterials \cite{el2021digital,el2022mechanical}, snap-through mechanisms \cite{mei2021mechanical}, electroactive actuators \cite{el2024intelligent}, and geometric nonlinearities of bistable spring-mass chains \cite{bilal2017bistable,ion2017digital}. In a separate domain, wave-based computing has also recently emerged, utilizing the rich dynamics of wave scattering and dispersion for the same purpose \cite{zangeneh2021analogue}. Computational functions such as spatial derivatives for image detection and other complex tasks have been demonstrated across acoustic and optical systems \cite{silva2014performing,pors2015analog,cordaro2019high,zuo2018acoustic,ren2021smp}. Inspired by this progress and the surge in machine learning, research efforts tackling physical neural networks \cite{lin2018all,weng2020meta,luo2020probability,zeng2025high}, reservoir computers \cite{shougat2021hopf}, and other variations of physical computing systems with learning and adaptation capabilities \cite{lee2022mechanical} have also gained momentum. One such example employs neuromorphic metasurfaces where trained subwavelength layers of resonant unit cells manipulate incident waves in a fashion similar to the interaction between neurons in artificial neural networks (ANNs) \cite{wu2020neuromorphic}. This analogy enables a class of intelligent structures that perform complex classification tasks \cite{moghaddaszadeh2024mechanical}. While effective, the overarching constraints on wave propagation, whether elastic, acoustic or electromagnetic, including reciprocity, transmission symmetry, and static dispersion patterns, have inherently limited these systems to single-task configurations. As such, their inability to re-adapt to new information or concurrently perform multiple tasks (i.e., compute in parallel) still represents a major hindrance to advancing systems with faster and broader physical computing capabilities. 

The approach presented here leverages the multiplexing effect achieved through the temporal modulation of the constitutive properties of the building blocks (unit cells) that form a computing metasurface \cite{mousa2024parallel}. Such modulation enables monochromatic incident waves to self-generate additional harmonics upon being transmitted through the modulated layers. These harmonics exhibit spectra that are distinct from the fundamental input \cite{shaltout2019spatiotemporal,sapienza2023apl, adlakha2020frequency}, allowing the notion of parallel computing through different frequency lanes to be created. As illustrated in Fig.~\ref{Fig1}a, in their conventional forms, digital classifiers are employed in various forms, e.g., binary classifiers, which distinguish between two classes; multiclass classifiers which categorize inputs into one of several classes, and multilabel classifiers which can assign that labels to a single input. The present study aims to develop a neuromorphic metasurface capable of several (independent) multiclass classifications within a single physical system, namely an elastic substrate that consists of a uniform material and metasurface layers positioned at two locations (Fig.~\ref{Fig1}b). By modulating the two metasurface layers at different frequencies, a large set of up- and down-shifted harmonics are generated, yielding an abundance of wave propagation channels. While some of these harmonics are interdependent, the analytical framework shown later ensures the availability of entirely independent channels, allowing the implementation of multiple classifiers within a single system that interprets the features of different inputs and reflects their classification at a prescribed readout terminal. The theory behind the time-modulated unit cell, waveguide array design, and the overall architecture of the neuromorphic system and its theory of operation are explained in detail, followed by a discussion of the results and the system's performance. The proposed approach represents a notable advancement in the field of mechanical computing and neuromorphic systems.

\section{Time-modulated neuromorphic metasurface}
\label{time_modulated_neuromorphic_metasurface}
\subsection{Operational principle}

\begin{figure}[t!]
\centering
\includegraphics[width=\linewidth]{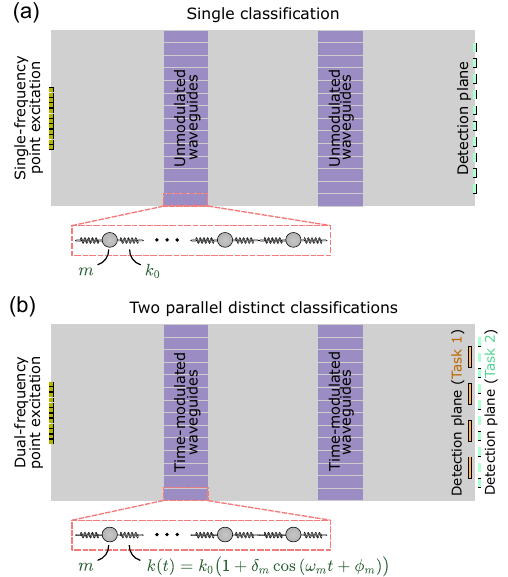}
\caption{(a) A single-task neuromorphic metasurface with two metasurface layers comprising waveguides of unmodulated unit cells, each with a lumped mass $m$ and a constant time-invariant stiffness $k_0$. (b) A multi-task neuromorphic metasurface with two metasurface layers comprising waveguides of time-modulated unit cells, each with a lumped mass $m$ and a time-varying stiffness $k(t) = k_{0} \big(1+\delta_m \cos{(\omega_m t + \phi_m)}\big)$. The left side of the system is excited with two sets of features at distinct frequencies, while the right side acts as a detection plane, indicating the outcomes of the different computations carried out via wave energy focusing at different frequencies corresponding to the different tasks.}
\label{Fig2}
\end{figure}

A neuromorphic metasurface is a physical form of a neural network that can perform artificial intelligence tasks (e.g., supervised learning and classification) via wave scattering. It is comprised of several optical \cite{wu2020neuromorphic, leonard2021co} or acoustic \cite{lin2022anomalous} metasurface layers that are sandwiched between uniform substrates, and manipulate incident waves to perform complex computational tasks. The computational input to a neuromorphic metasurface comes from the outside environment through pressure, thermal, or dynamic loads. For example, the features of a classification dataset can be coded into the actuation amplitudes of a set of sources (vibrational or acoustic drivers), which simulate external excitations. These excitations generate waves that propagate through multiple layers of metasurfaces composed of tunable waveguides. By locally altering the magnitude and phase of the transmitted wave at each waveguide (through training), the propagating wavefront can be made to focus the bulk amount of energy at a specific spatial coordinate within the detection plane at one end of the system, which is how the system conducts a ``single" multiclass classification (see Fig.~\ref{Fig2}a). While such systems can reliably classify different inputs from the same dataset, shifting to a new problem requires either the construction of a new set of metasurface layers or a geometric reconfiguration of the current ones \cite{moghaddaszadeh2024mechanical}. More importantly, conducting several such problems in parallel requires stacking multiple units, which remarkably increases the footprint of the physical computing system, raising questions about practicality, size, and cost-efficiency. 

To overcome this, the approach presented here invokes the emerging harmonics in time-variant waveguides whose stiffness profiles are temporally modulated via a prescribed speed, phase, and amplitude. When adequately tuned, the neuromorphic metasurface's response within these newly generated frequencies can be utilized to carry out new tasks, independent of one another, allowing different computational outcomes corresponding to different problems to be concurrently read out at the same detection plane, albeit at different frequencies (see Fig.~\ref{Fig2}b). The aforementioned concept first necessitates a redesign of the base unit cell, replacing the elastically static waveguide used in single-task metasurfaces with a time-varying one. While such time variations can be realized in practice via rotations \cite{attarzadeh2020experimental}, piezoelectricity \cite{marconi2020experimental}, or basic circuitry \cite{moghaddaszadeh2021nonreciprocal}, the next few subsections focus on the analytical derivation of the scattering matrix and transmission coefficients of such time-modulated waveguides, which is critical to the development of a dual classifier neuromorphic metasurface, as will be shown in Sec.~\ref{section_metasurface}.

\subsection{Transfer matrix of a time-modulated unit cell}
\label{section_transfer_matrix}
\begin{figure}[t!]
\centering
\includegraphics[width=\linewidth]{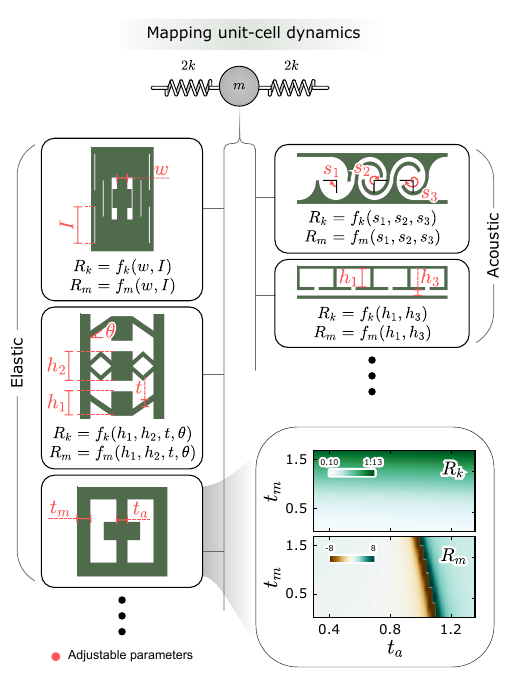}
\caption{Examples of solid elastoacoustic unit cells, which can be represented by a generalizable spring-mass model \cite{lee2018mass,lee2023mode,zuo2017mathematical,li2015metascreen}. The corresponding unit cell's effective inertial and elastic indicators $R_m$ and $R_k$ can be derived directly from the unit cell’s constitutive parameters, as illustrated in the bottom-right example, where the unit cell shown at the bottom left is expanded into an array of six unit cells, and the equivalent lumped model is utilized to generate $R_m$ and $R_k$ \cite{moghaddaszadeh2024mechanical}. Once the pertinent dynamics are mapped from the geometric features of the solid cells, the framework derived in Sec.~\ref{time_modulated_neuromorphic_metasurface} can be applied to any of the above configurations.}
\label{Fig3}
\end{figure}

Consider the single spring-mass cell consisting of a lumped mass $m$ and a time-dependent stiffness $k(t)$ of the form
\begin{equation}
\label{modulation}
    k(t) = k_{0} \big(1+\delta_m \cos{(\omega_m t + \phi_m)}\big)
\end{equation}
where $k_{0}$ is the static stiffness while $\delta_m$, $\omega_m$ and $\phi_m$ denote the modulation depth, frequency, and phase, respectively. The cell's motion equations can be derived as follows:
\begin{subequations}
\label{EOM}
\begin{align}
    m \Ddot{u}_2 &= 2k (u_1 + u_3 - 2u_2)
    \\
    f_{1} &= 2k (u_2 - u_1)
    \\
    f_{3} &= 2k (u_3 - u_2)
\end{align}
\end{subequations} 
where $u_{1,3}$ and $f_{1,3}$ represent the displacement and forcing at the unit-cell boundaries, whereas $u_2$ is the displacement of the lumped mass. Assuming a harmonic input of frequency $\omega$, owing to the time-dependence of the stiffness, the system responds at the driving frequency $\omega$ and at secondary harmonics given by $\omega + \alpha \omega_m$, such that
\begin{subequations}
\label{sol}
\begin{align}
    u_i &=  \sum^{\infty}_{\alpha=-\infty} \hat{u}_i^{[\omega + \alpha \omega_m]} e^{\ii (\omega + \alpha \omega_m t)}; \hspace{1cm} i=1,2,3
    \\
    f_i &=  \sum^{\infty}_{\alpha=-\infty} \hat{f}_i^{[\omega + \alpha \omega_m]} e^{\ii (\omega + \alpha \omega_m t)}; \hspace{1cm} i=1,3
\end{align}
\end{subequations}
where $\ii=\sqrt{-1}$ and the superscript $[\omega^{\ast}]$ in Eq.~(\ref{sol}) represents the variable of interest corresponding to a wave of frequency $\omega^{\ast}$ (henceforth referred to as the frequency channel $\omega^{\ast}$). For example, $\smash{\hat{u}_i^{[\omega + \alpha \omega_m]}}$ represents the displacement amplitude at the $\omega + \alpha \omega_m$ channel. Substituting Eqs.~(\ref{sol}) and (\ref{modulation}) into Eq.~(\ref{EOM}), considering harmonics up to the first order (i.e., $\alpha=[-1,0,1]$), and ignoring higher-order $\delta$ terms (i.e., $\delta^2$, $\delta^3$, etc), a transfer matrix $\mathbf{T}$ can be extracted that directly relates the forces and displacements on one end of the cell to the other within the three frequency channels $\omega$, $\omega^- = \omega - \omega_m$, and $\omega^+ = \omega + \omega_m$, as follows:
\begin{equation}
    \label{tf}
    \mathbf{\tilde{u}}_3 = \mathbf{T}~\mathbf{\tilde{u}}_1
\end{equation}
where $\mathbf{\tilde{u}}_{1,3} = [\hat{u}_{1,3}^{[\omega^-]} \hspace{0.2cm} \hat{f}_{1,3}^{[\omega^-]} \hspace{0.2cm} \hat{u}_{1,3}^{[\omega]} \hspace{0.2cm} \hat{f}_{1,3}^{[\omega]} \hspace{0.2cm} \hat{u}_{1,3}^{[\omega^+]} \hspace{0.2cm} \hat{f}_{1,3}^{[\omega^+]}]^T$ (see Appendix for detailed structure of $\mathbf{T}$). One or more of these frequency channels can be utilized to host independent intelligence tasks via signal multiplexing, as will be shown later, thus manifesting the concept of parallel computing in a single mechanical substrate. Second- and higher-order harmonics can be likewise utilized as additional computational labels, although these are more practically challenging since they comprise less energy \cite{moghaddaszadeh2021nonreciprocal}. The transfer matrix in Eq.~(\ref{tf}) can be used to derive the scattering properties of a periodic array of such cells, as will be shown next. To enable more than two independent computations to be conducted, more than two unit-cell parameters must be modulated, which is feasible in unit-cell geometries with more intricate details or internal architectures (e.g., multiple resonators, higher-order lattices, etc), thereby giving rise to new independent computational channels.

It is worth noting that we adopt a lumped parameter model for the unit cell with effective inertial and elastic indicators $R_m$ and $R_k$ that can be matched with several configurations of solid continua, so as not to limit the developed framework to a single specific metamaterial design. Moreover, the same modeling approach can also be extended to other physical domains, such as optical and acoustic ones. For instance, the unit cells depicted in Fig.~\ref{Fig3} represent examples of elastic and acoustic unit cells. Each of these unit cells can be modeled as spring-mass systems whose constitutive parameters can then be used to obtain the transmission maps, a requisite step for the neuromorphic metasurface design, as will be outlined next.
\subsection{Analytical model of a single waveguide}
\label{section_waveguide}
\begin{figure*}[t!]
\centering
\includegraphics[width=\linewidth]{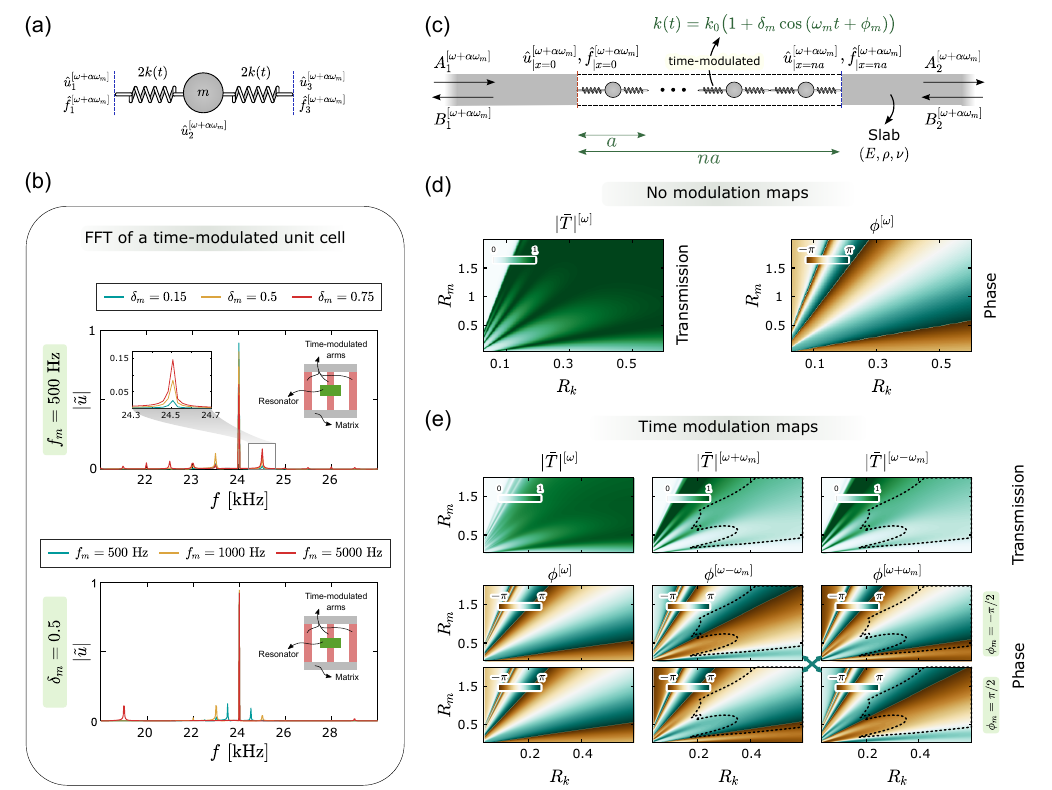}
\caption{Multifrequency transmission of time-modulated waveguides. (a) General design of a unit cell, the building block of the metasurface, where $\smash{\hat{u}_i^{[\omega + \alpha \omega_m]}}$ and $\smash{\hat{f}_i^{[\omega + \alpha \omega_m]}}$ represent displacement and forcing at a frequency $\omega + \alpha \omega_m$. (b) Frequency stability analysis of a select time-modulated unit cell performed at various modulation amplitudes (top) and modulation frequencies (bottom), showing that a careful selection of the operating and modulation frequencies is required to ensure the stability of the harmonics and avoid spectral leakage to neighboring frequencies. Here, FFT denotes fast Fourier transform. (c) A single waveguide made of an array of $n$ unit cells, where $\smash{A_{1,2}^{[\omega + \alpha \omega_m]}}$ and $\smash{B_{1,2}^{[\omega + \alpha \omega_m]}}$ denote the amplitudes of forward and backward propagating waves on either side of the waveguide for the frequency channel $\omega+\alpha \omega_m$. $a$ is the lattice constant and $\alpha$ can be $-1$, $0$, or $1$. (d) Transmission amplitude $|\bar{T}|$ and phase $\phi$ of a waveguide made of $n=6$ unmodulated cells with stiffness $k(t) = k_0$ for an incident wave with a frequency $\omega=150$ kHz. Here, $R_m$ and $R_k$ are the effective mass and stiffness ratios as defined in the text. (e) Transmission amplitude $|\bar{T}|$ and phase $\phi$ for the three transmitted harmonics from a waveguide made of $n=6$ time-modulated cells with the following modulation parameters: $\delta = 0.15$, $\omega_m = 7$ kHz, $\phi_m = -\pi/2$ and $\pi/2$, and stiffness $k(t) = k_{0} \big(1+\delta_m \cos{(\omega_m t + \phi_m)}\big)$, generated from an incident wave with frequency $\omega=150$ kHz. Bordered regions highlight the design space where $\smash{0.8<\bar{T}^{[\omega]}<1}$ and $\smash{0.1<|\bar{T}^{[\omega^{\pm}]}|<0.3}$, ensuring acceptable transmission while maintaining $2\pi$ phase tunability.}
\label{Fig4}
\end{figure*}

Consider a single-dimensional array of $n$ cells, each having a length $a$, forming a waveguide of length $na$, as shown in Fig.~\ref{Fig4}c. The displacement field at both sides of the array can be defined as
\begin{equation}
\label{wave_prop}
        u_{l,r}^{[\omega^{\ast}]} = \left( A_{1,2}^{[\omega^{\ast}]} e^{-\ii \tilde{\kappa}^{[\omega^{\ast}]} x_{l,r}} + B_{1,2}^{[\omega^{\ast}]} e^{\ii \tilde{\kappa}^{[\omega^{\ast}]} x_{l,r}} \right) e^{\ii \omega^{\ast} t}
\end{equation}

where the subscripts $l$ and $r$ refer to the left and right ends of the array, respectively, while $x_l=x$ and $x_r=x-na$. $\smash{A_{1,2}^{[\omega^{\ast}]}}$ and $\smash{B_{1,2}^{[\omega^{\ast}]}}$ denote the amplitudes on either side of the waveguide for the frequency channel $\omega^{\ast}$ (i.e., $\omega^-$, $\omega$, or $\omega^+$). Note that $\smash{\tilde{\kappa}^{[\omega^{\ast}]}}$ represents the wavenumber of the frequency channel $\omega^{\ast}$. The forces associated with forward and backward propagating waves are equal to $f^{[\omega^{\ast}]}=\mp\ii \omega^{\ast} z u$, with $z = \rho A c$ being the mechanical impedance and $c = \sqrt{ E / [\rho(1-\nu^2)]}$ representing the longitudinal wave speed, where $E$, $\rho$, $\nu$, and $A$ are the elastic modulus, density, Poisson's ratio, and cross-sectional area of the slab. As a result, Eq.~(\ref{wave_prop}) can be cast in matrix form as follows:
\begin{equation}
\label{pre_M}
    \begin{bmatrix}
        \hat{u}_{l,r}^{[\omega^{\ast}]}\\[0.2cm]
        \hat{f}_{l,r}^{[\omega^{\ast}]}
    \end{bmatrix}
    = 
    \begin{bmatrix}
        e^{-\ii \tilde{\kappa}^{[\omega^{\ast}]} x_{l,r}} & e^{\ii \tilde{\kappa}^{[\omega^{\ast}]} x_{l,r}} \\[0.2cm]
        -\ii \omega^{\ast} z e^{-\ii \tilde{\kappa}^{[\omega^{\ast}]} x_{l,r}} & \ii \omega^{\ast} z e^{\ii \tilde{\kappa}^{[\omega^{\ast}]} x_{l,r}}
    \end{bmatrix}
    \begin{bmatrix}
        A_{1,2}^{[\omega^{\ast}]} \\[0.2cm]
        B_{1,2}^{[\omega^{\ast}]}
    \end{bmatrix}
\end{equation}
Right at the boundaries where $x = 0$ and $x = na$, Eq.~(\ref{pre_M}) simplifies to
\begin{equation}
    \begin{bmatrix}
        \hat{u}_{l,r}^{[\omega^{\ast}]}\\[0.2cm]
        \hat{f}_{l,r}^{[\omega^{\ast}]}
    \end{bmatrix}
    = \mathbf{M}^{[\omega^{\ast}]}
    \begin{bmatrix}
        A_{1,2}^{[\omega^{\ast}]} \\[0.2cm]
        B_{1,2}^{[\omega^{\ast}]}
    \end{bmatrix} \hspace{1cm}
\end{equation}
where     
\begin{equation}
    \mathbf{M}^{[\omega^{\ast}]} =
    \begin{bmatrix}
        1 & 1        \\ 
        -\ii (\omega^{\ast}) z  & \ii (\omega^{\ast}) z
    \end{bmatrix}
\end{equation}

Since only $\alpha$ values of $-1$, $0$, and $1$ are considered, a total matrix $\mathbf{M}_t$ assembled from the individual $\mathbf{M}^{[\omega^{\ast}]}$ matrices for the three $\alpha$ values can be defined as
\begin{equation}
    \mathbf{M}_t = 
    \begin{bmatrix}
        \mathbf{M}^{[\omega^-]} & \mathbf{0} & \mathbf{0}\\
        \mathbf{0} & \mathbf{M}^{[\omega]} & \mathbf{0}\\
        \mathbf{0} & \mathbf{0} & \mathbf{M}^{[\omega^+]}
    \end{bmatrix}
\end{equation}
where $\mathbf{0}$ is a $2\times2$ zero matrix. Following this, we arrive at $\mathbf{\tilde{u}}_{l,r} = \mathbf{M}_t  \mathbf{c}_{1,2}$, where $\smash{\mathbf{c}_{1,2} = [A_{1,2}^{[\omega^-]}\hspace{0.2cm} B_{1,2}^{[\omega^-]}\hspace{0.2cm} A_{1,2}^{[\omega]}\hspace{0.2cm} B_{1,2}^{[\omega]}\hspace{0.2cm} A_{1,2}^{[\omega^+]}\hspace{0.2cm} B_{1,2}^{[\omega^+]}]^T}$. At the same time, the relation $\mathbf{\tilde{u}}_{l} = \mathbf{T}^r \mathbf{\tilde{u}}_{r}$ describes the global transfer matrix $\mathbf{T}^r$ relating the two sides of the array. By combining the two previous equations together, the connection $\mathbf{c}_2 = \mathbf{S} \mathbf{c}_1$ can be made where $\mathbf{S}=\mathbf{M}_t^{-1}\mathbf{T}^{r}\mathbf{M}_t$ defines the array's scattering matrix. Assuming no reflection from the right side (i.e., $\smash{B_2^{[\omega + \alpha \omega_m]}=0}$ for $\alpha = [-1,0,1]$) and the incident wave to only have a fundamental component (i.e., $\smash{A_1^{[\omega + \alpha \omega_m]}=0}$ for $\alpha = [-1,1]$), the transmission coefficients can be defined as the amplitude ratio of the transmitted wave to the incident one: $\smash{\bar{T}^{[\omega + \alpha \omega_m]}=A_2^{[\omega + \alpha \omega_m]}/A_1^{[\omega]}}$.

It is important to note that since this work considers only longitudinal wave excitation from the point sources, the excitation frequencies and the design of the plate hosting the neuromorphic metasurface are chosen such that the frequency–thickness product is well below $1$ MHz$\,$mm, providing the condition needed for only fundamental Lamb wave modes to propagate \cite{rose2014ultrasonic,li2019modelling,li2024mode}. While some degree of mode conversion to shear waves is inherent to any nonideal sources, an elastic metasurface can be specifically engineered to maintain the primary in-plane mode as the dominant wave propagating through the structure, as demonstrated in related studies \cite{lee2023mode,lee2018mass}.

To understand modulation effects at the three harmonics, consider a waveguide consisting of $n=6$ unit cells of length $a=3$ mm attached to aluminum slabs ($E=70$ GPa, $\rho=2700$ kg/m$^3$, and $\nu=0.33$) that are $3$ mm wide and $1$ mm thick, as shown in Fig.~\ref{Fig4}c. The effective mass and stiffness ratios, $R_m = m / \Bar{m}$ and $R_k = k_0 /  \Bar{k}$, can be defined such that $\Bar{m} = \rho A a$ and $\Bar{k} = E A / [a(1-\nu^2)]$ represent the mass and axial stiffness, respectively, of an aluminum plate of the same size. The transmission amplitude $|\bar{T}|$ and phase $\phi = \angle \bar{T}$ of a single waveguide in the absence of modulation [i.e., $k(t) = k_0$] are shown in Fig.~\ref{Fig4}d for an input wave of $\omega=150$ kHz. Upon setting the modulation parameters $\delta = 0.15$, $\omega_m = 7$ kHz, and $\phi_m = -\pi/2$, the transmission amplitude and phase at the three harmonic frequencies (for the same input wave) are shown in Fig.~\ref{Fig4}e. Comparing the two figures, it can first be observed that a portion of the energy is transferred from the fundamental channel $\omega$ to the secondary $\omega^+$ and $\omega^-$ channels as a direct result of the modulation. On the other hand, while the phase profile at the fundamental channel, i.e., $\phi^{[\omega]}$, remains practically unchanged in the absence and presence of modulation, the secondary phases $\phi^{[\omega^-]}$ and $\phi^{[\omega^+]}$ appear to have their own unique profiles. To further investigate this, the phase profiles are reexamined when $\phi_m = \pi/2$ (while everything else is kept the same), as illustrated in the bottom panel of Fig.~\ref{Fig4}e. As expected, $\phi^{[\omega]}$ remains unchanged, which confirms the independence of the fundamental frequency channel and its indifference to changes made to the modulation phase $\phi_m$. We also note that the new  $\phi^{[\omega^+]}$ and $\phi^{[\omega^-]}$ profiles shift by $\pi$ and $-\pi$, respectively, as a result of $\phi_m$ changing from $-\pi/2$ to $\pi/2$.

These results demonstrate a level of codependence between $\smash{\phi^{[\omega^+]}}$ and $\smash{\phi^{[\omega^-]}}$. Specifically, increasing the modulation phase $\phi_m$ by a $\pi$ amount adds $\pi$ to the $\smash{\phi^{[\omega^+]}}$ profile and subtracts $\pi$ from its $\smash{\phi^{[\omega^-]}}$ counterpart. However, both profiles remain independent of $\phi^{[\omega]}$, allowing either channel to ``host" an independent task. The use of $\phi_m$ as an additional degree of freedom in the design and training of the time-modulated neuromorphic metasurface will be detailed next. Finally, the bordered regions, outlined by dashed lines in Fig.~\ref{Fig4}e, highlight the design space in which $\smash{0.8<|\bar{T}^{[\omega]}|<1}$ and $\smash{0.1<|\bar{T}^{[\omega^{\pm}]}|<0.3}$ for each case. Within this region, the modulated waveguide retains a consistent transmission amplitude while providing complete phase tunability over a $2 \pi$ range---a criterion that is critical for the successful operation of the wave-based dual classifier. 

The choice of modulation parameters, i.e., modulation amplitude and frequency, strongly influences the transmission ranges of the frequency channels and the degree of spectral leakage into neighboring frequencies. To examine this, an analytical study is conducted to investigate the frequency stability of a time-modulated unit cell. For the purpose of this analysis, we select an actual solid unit cell, depicted in Fig.~\ref{Fig4}b, which is comprised of an aluminum matrix, a brass resonator, and two slender aluminum beams (resonator arms) that attach the resonator to the matrix, similar to the configuration used in Ref.~\cite{moghaddaszadeh2024mechanical}. An array of six unit cells is integrated into an aluminum waveguide, with waves excited from one end. The stiffness of each unit cell is time modulated at various modulation frequencies and amplitudes. It is shown that increasing the modulation frequency leads to a broader spectral spread, resulting in overlap (crosstalk) between computational channels, which is detrimental to the computing mechanism. As such, the operating frequencies of the computation channels must be carefully selected to avoid any overlap. The results indicate that keeping the modulation frequency below approximately $10\%$ of the fundamental excitation frequency minimizes spectral leakage, whereas higher modulation frequencies can destabilize the system and redistribute energy away from the desired frequency components.

\section{Dual classifier neuromorphic metasurface}
\label{section_metasurface}
\begin{figure*}[t!]
\centering
\includegraphics[width=\linewidth]{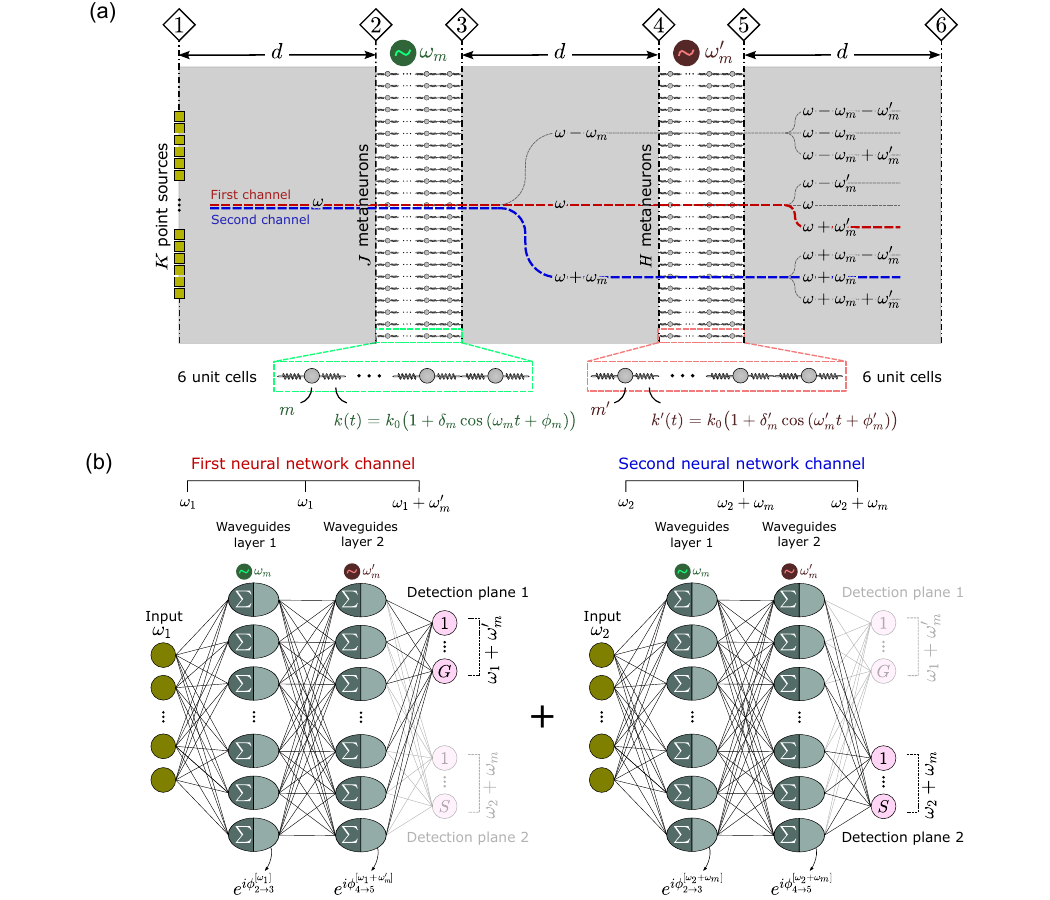}
\caption{(a) The proposed time-modulated neuromorphic metasurface is composed of an input layer (axis $1$), two time-modulated metasurface layers (spanning axes $2 \rightarrow 3$ and $4 \rightarrow 5$) , and a detection plane (axis $6$). Each two consecutive layers are displaced by a distance $d$. The metasurface layers are comprised of the waveguides illustrated in Fig.~\ref{Fig4}c. The frequency components propagating in each segment, including the excitation frequency and all harmonics generated by the time-modulated metasurfaces, are listed. (b) Parallel classification logic diagrams. The first neural network channel corresponding to the frequency path $\omega_1 \rightarrow \omega_1 \rightarrow \omega_1 + \omega^{\prime}_m$ is shown on the left with $\smash{\phi^{[\omega_1]}_{2\rightarrow3,j}}$ and $\smash{\phi^{[\omega_1 + \omega^{\prime}_m]}_{4\rightarrow5,h}}$ as trainable parameters, while the second neural network channel corresponding to the frequency path $\omega_2 \rightarrow \omega_2 + \omega_m \rightarrow \omega_2 + \omega_m$ is shown on the right with $\smash{\phi^{[\omega_2 + \omega_m]}_{2\rightarrow3,j}}$ and $\smash{\phi^{[\omega_2 + \omega_m]}_{4\rightarrow5,h}}$ as trainable parameters.}
\label{Fig5}
\end{figure*}

\subsection{Scattering channels}
\label{subsection_metasurface_scattering}

A detailed outline of the dual classifier neuromorphic metasurface is provided in Fig.~\ref{Fig5}a. It consists of an input layer (labeled as axis $1$), two time-modulated metasurface layers, and a detection plane (axis $6$). The first metasurface layer is positioned between axes $2$ and $3$, and contains $J = 500$ waveguides, which are comprised of mass $m$ and a time-modulated stiffness $k(t) = k_{0} \big(1+\delta_m \cos{(\omega_m t + \phi_m)}\big)$. The second metasurface layer is positioned between axes $4$ and $5$, and contains $H = 500$ waveguides which are comprised of $m^{\prime}$ and $k^{\prime}(t) = k_{0} \big(1+\delta^{\prime}_m \cos{(\omega^{\prime}_m t + \phi^{\prime}_m)}\big)$. All waveguide parameters match those used in Sec.~\ref{section_waveguide}. It is worth noting that an individual waveguide can also be referred to as a ``metaneuron'' since it represents the physical (metamaterial) equivalent of a neuron in an ANN. The phase tunability of each waveguide within a metasurface layer is similar to the trainability required in an ANN. Depending on the problem at hand, $K$ features of each sample in the input dataset are provided as left-end point excitations (i.e., monopoles). Finally, we emphasize that while all the aforementioned parameters are specific for the examples showcased in Sec.~\ref{subsection_metasurface_results}, they are chosen without loss of generality and can be readily adjusted to accommodate different problems or physical constraints.

Assuming circular waves emitting from each point source, the $x$ displacement at the entrance to each waveguide $j$ at axis $2$ can be derived as \cite{giurgiutiu2007structural}
\begin{equation}
\label{axis1}
     u_{2,j}= 
     \sum^{K}_{k=1} \frac{u_{1,k}}{\sqrt{|\mathbf{p}_{2,j}- \mathbf{p}_{1,k}|}}e^{\ii\big(\omega t - \tilde{\kappa} ({|\mathbf{p}_{2,j}- \mathbf{p}_{1,k}|})\big)}
\end{equation}
where $u_{1,k}$ is the amplitude of the $k^{\text{th}}$ point source at axis $1$, and $\mathbf{p}_{2,j}$ and $\mathbf{p}_{1,k}$ denote the position vectors of the $j^{\text{th}}$ waveguide and the $k^{\text{th}}$ point source, respectively. As derived in Sec.~\ref{section_waveguide}, the wavefield exiting the first layer of time-modulated waveguides contains three distinct frequencies: the fundamental frequency $\omega$ matching the input excitation, in addition to the down- and up-shifted harmonics, $\omega-\omega_m$ and $\omega+\omega_m$, respectively, which are functions of the modulation frequency. A time-modulated waveguide can therefore be perceived as a black box, whose influence on the impinging signal is captured by a complex exponential resulting from the applied phase shift, coupled with the corresponding transmission coefficient:
\begin{equation}
\label{axis3}
     u_{3,j}^{[\omega^{\ast}]}= |\bar{T}|_{2\rightarrow3,j}^{[\omega^{\ast}]} e^{\ii \phi^{[\omega^{\ast}]}_{2\rightarrow3,j}} u_{2,j}
\end{equation}
where $u_{3,j}^{[\omega^{\ast}]}$ is the $x$ displacement at the output of waveguide $j$ at axis $3$ for three aforementioned frequencies. Moreover, $\smash{|\bar{T}|_{2\rightarrow3,j}^{[\omega^{\ast}]}}$ and  $\smash{\phi^{[\omega^{\ast}]}_{2\rightarrow3,j}}$ are the transmission amplitude and phase imposed by the $j^{\text{th}}$ waveguide between axes $2$ and $3$. While $\phi^{[\omega]}_{2\rightarrow3,j}$ depends only on the static parameters $m$ and $k_0$, $\smash{\phi^{[\omega-\omega_m]}_{2\rightarrow3,j}}$ and $\smash{\phi^{[\omega+\omega_m]}_{2\rightarrow3,j}}$ are influenced (and tuned) by the first metasurface layer modulation phase $\phi_m$, as discussed in Sec.~\ref{section_waveguide} and illustrated in Fig.~\ref{Fig4}c. The harmonics of Eq.~(\ref{axis3}) can be considered new point sources based on the Huygens-Fresnel principle \cite{goodman2005}, giving rise to three waves that propagate from axis $3$ to axis $4$, as follows:
\begin{equation}
\label{axis4}
     u_{4,h}^{[\omega^{\ast}]}= \sum^{J}_{j=1} \frac{u_{3,j}^{[\omega^{\ast}]}}{\sqrt{|\mathbf{p}_{4,h}- \mathbf{p}_{3,j}|}}e^{\ii \tilde{\kappa}^{[\omega^{\ast}]} ({|\mathbf{p}_{4,h}- \mathbf{p}_{3,j}|})}
\end{equation}
with the notation following the same pattern defined earlier. As can be inferred from the foregoing analysis, the second layer of time-modulated waveguides imparts the same harmonic conversions on each of the three incident frequencies, yielding a set of nine propagating waves corresponding to $\omega - \omega_m - \omega^{\prime}_m$, $\omega - \omega_m$, $\omega - \omega_m + \omega^{\prime}_m$, $\omega - \omega^{\prime}_m$, $\omega$, $\omega + \omega^{\prime}_m$, $\omega + \omega_m - \omega^{\prime}_m$, $\omega + \omega_m$, and $\omega + \omega_m + \omega^{\prime}_m$. Expressions for $\smash{u_{5,h}^{[\omega^{\ast}]}}$ and $\smash{u_{6,o}^{[\omega^{\ast}]}}$ can be obtained in a manner similar to Eqs.~(\ref{axis3}) and (\ref{axis4}), respectively, where $h$ and $o$ denote the indices of points on axes $5$ and $6$, respectively. At axis $6$, the detection plane is split into small equal-length sections corresponding to the classes of the dataset of interest.

\subsection{Parallel classifications of two distinct datasets}
\label{subsection_metasurface_parallel_classifications}

The wave propagation pattern within each frequency channel is similar to a customized ANN architecture. A conventional ANN consists of multiple hidden layers with trainable weights and biases that connect the input, hidden, and output layers \cite{goodfellow2016deep}. At each layer, a typically fixed activation function is applied to regulate the output of hidden-layer neurons. Similarly, in the neuromorphic metasurface, consisting of input, metasurface, and output layers, wave scattering in each of the nine frequency channels can be manipulated by training (i.e., mechanically tuning) $\smash{\phi^{[\omega^{\ast}]}_{2\rightarrow3,j}}$ and $\smash{\phi^{[\omega^{\ast}]}_{4\rightarrow5,h}}$ in the first and second metasurface layers, respectively. However, unlike ANNs where weights are adjustable during training, the weights of a neuromorphic metasurface depend solely on the size and material properties of its different layers which, once set, cannot be changed. To address this, the aforementioned tunable phases play an role equivalent to activation functions and can be effectively tuned to accomplish the physical computation at hand.

Consider the schematic diagram shown on the left side of Fig.~\ref{Fig5}b representing the wave propagation within the first neural network. Input features of frequency $\omega_1$ scatter and sum up based on Eq.~(\ref{axis1}). The parameters of this equation are based solely on the geometric and material properties of the uniform (aluminum) substrate between axes $1$ and $2$ and, therefore, they represent a set of nontrainable hyperparameters in the neural architecture. The first metasurface layer, modulated at a frequency $\omega_m$, then manipulates the incident wavefront at axis $2$ in accordance with Eq.~(\ref{axis3}), resulting in the generation of three waves at the fundamental, down-shifted, and up-shifted harmonics of the input frequency $\omega_1$. Here, we utilize $\omega_1$ as the carrier frequency for the first neural network channel (see Fig.~\ref{Fig5}b), disregarding the other two harmonics. Accordingly, the trainable parameter for each metaneuron in the first metasurface layer is $\phi^{[\omega_1]}_{2\rightarrow3,j}$. At axis $3$, the resultant wavefront is treated as an array of new point sources from which waves scatter and sum in a manner analogous to the original input waves. Subsequently, the second metasurface layer, modulated at a frequency $\omega^{\prime}_m$, manipulates the waves incident at axis $4$, generating three new harmonics at axis $5$. The up-shifted harmonic $\omega_1 + \omega^{\prime}_m$ is chosen to resume the computational path from the moment the waves exit the second metasurface layer till they reach the detection plane, thereby defining the frequency path of the first classification task as $\omega_1$ $\rightarrow$  $\omega_1$ $\rightarrow$ $\omega_1 + \omega^{\prime}_m$ before, between, and after the two metasurface layers, respectively. This also implies that the trainable parameter for each metaneuron in the second metasurface layer is $\smash{\phi^{[\omega_1 + \omega^{\prime}_m]}_{4\rightarrow5,h}}$, as detailed in Sec.~\ref{subsection_metasurface_scattering}. Finally, waves at frequency $\omega_1 + \omega^{\prime}_m$ coalesce at the detection plane, focusing the bulk of their energy at the vertical location marking the correct classification label. To identify the selected class, the detection plane is partitioned into $G$ segments (labels of the first dataset), such that the segment sensing the highest intensity denotes the predicted class.

In a simultaneous manner, the features of a second dataset are fed to the neuromorphic metasurface as input excitations of a frequency $\omega_2$, and an independent frequency path $\omega_2$ $\rightarrow$  $\omega_2 + \omega_m$ $\rightarrow$ $\omega_2 + \omega_m$ is then chosen for the second classification task, which is being concurrently carried out using the same physical system. Based on this frequency route, the corresponding trainable parameters in the first and second metasurface layers within the second neural network (right side of Fig.~\ref{Fig5}b) are $\smash{\phi^{[\omega_2 + \omega_m]}_{2\rightarrow3,j}}$ and $\smash{\phi^{[\omega_2 + \omega_m]}_{4\rightarrow5,h}}$, respectively. As with the first task, the detection plane of this channel is partitioned into $S$ segments, each representing an output label for the second dataset.

It is important to note that the ability to perform two distinct classifications in parallel relies on the independence of their trainable parameters, which is enabled by the selection of the right combination of frequency channels along which the two neural networks operate. For instance, in the first neural network, only the frequency channel $\omega_1$ is considered after the first metasurface layer. Consequently, the value of $\smash{\phi^{[\omega_1]}_{2\rightarrow3,j}}$ can be independently tuned via the unit-cell parameters, i.e., $k_0$ and $m$ of the $\smash{j^{\text{th}}}$ waveguide in the first metasurface layer, using the ``no modulation'' maps of Fig.~\ref{Fig4}d. On the other hand, for the second neural network, $\omega_2 + \omega_m$ represents the carrier frequency for the second task at the location where the waves exit the first metasurface layer, and therefore, $\smash{\phi^{[\omega_2 + \omega_m]}_{2\rightarrow3,j}}$ is used to determine the modulation phase $\phi_m$ (assuming preselected values of $\omega_m$ and $\delta$) for the $\smash{j^{\text{th}}}$ waveguide in the same first metasurface layer, using the ``time modulation'' maps of Fig.~\ref{Fig4}e. Following the same rationale, $\smash{\phi^{[\omega_2 + \omega_m]}_{4\rightarrow5,h}}$ is used to design the unit cell, i.e., determine $k^{\prime}_0$ and $m^{\prime}$, of the $l^{\text{th}}$ waveguide in the second metasurface layer, while $\smash{\phi^{[\omega_1 + \omega^{\prime}_m]}_{4\rightarrow5,h}}$ dictates the modulation phase $\phi^{\prime}_m$. This approach ensures the realization of two fully decoupled and independently trainable neural network channels within a single elastic medium.

\subsection{Task training and design}
\label{subsection_metasurface_training_Design}
\begin{figure*}[t!]
\centering
\includegraphics[width=\linewidth]{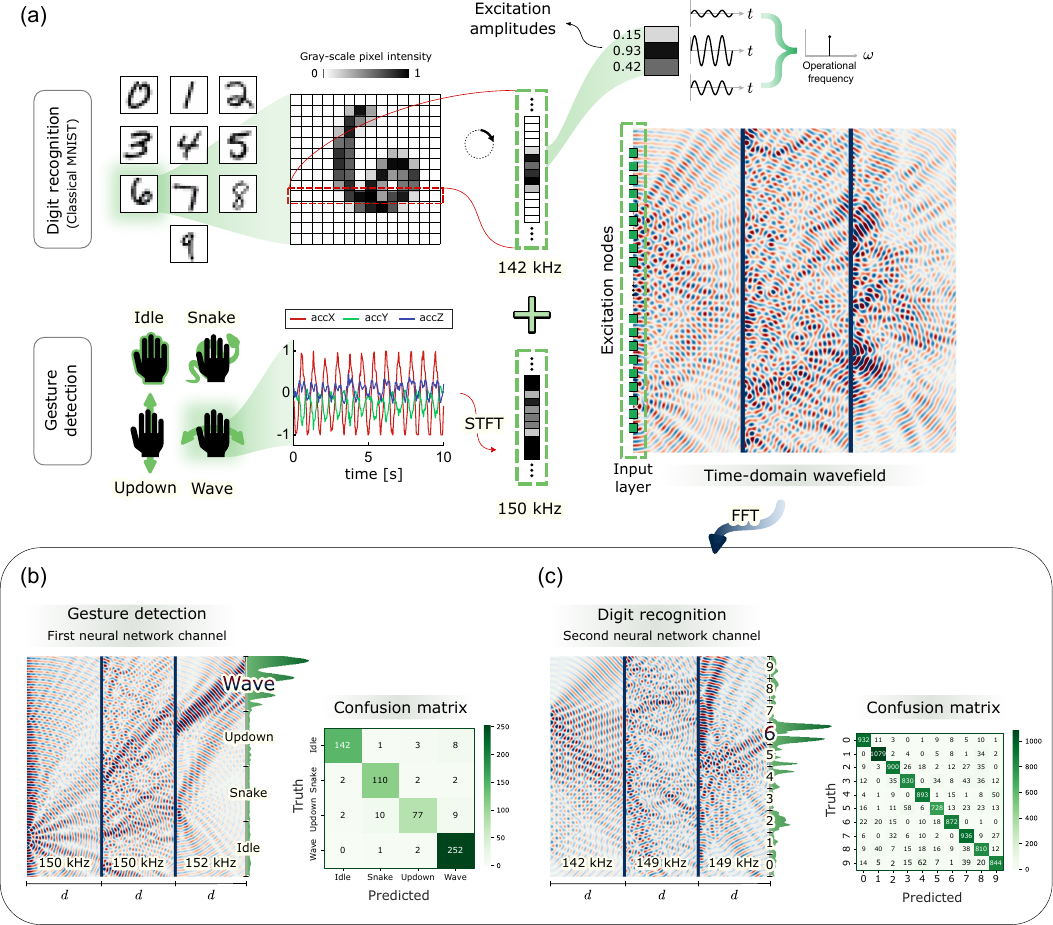}
\caption{Performance of the dual classifier neuromorphic metasurface. (a) The left-hand top and bottom panels depict the preprocessing of a single-digit sample ``6'' from the MNIST dataset and a gesture sample ``Wave'' from the gesture dataset, respectively, before they are introduced to the neuromorphic metasurface. The right-hand panel displays the aggregate wave scattering associated with all the harmonics generated by the neural network channels of both classifiers in the time domain. Here, STFT denotes short-time Fourier transform. (b),(c) Specific frequency-domain wavefields for two test samples along the first ($150\; \mathrm{kHz} \rightarrow 150\; \mathrm{kHz} \rightarrow 152\; \mathrm{kHz}$) and second ($142\; \mathrm{kHz} \rightarrow 149\; \mathrm{kHz} \rightarrow 149\; \mathrm{kHz}$) frequency paths corresponding to the gesture detection and MNIST digit recognition classifiers, respectively. The distribution of wave intensities at the detection planes of both wavefields is shown alongside the contour plots, signaling the high accuracy of both classifiers. This is further confirmed by the shown confusion matrices obtained from test samples for both problems. The wave propagation amplitudes in each of the wavefield segments separated by metasurface layers are individually normalized with respect to the same segment for better visualization.}
\label{Fig6}
\end{figure*}

A neuromorphic metasurface based on this framework is trained to concurrently execute two independent classifications for two different problems, namely a gesture (continuous motion) detection problem \cite{hymel2022edge} and the well-known MNIST digit recognition problem \cite{deng2012mnist}. The examples shown here are meant to demonstrate the capabilities of the proposed system with the understanding that the datasets themselves can be changed for varying applications. The gesture detection dataset is obtained via $15$ minutes of data collection using a MEMS accelerometer. A time history of $x$, $y$, and $z$accelerations is recorded with a sampling frequency of $62.5$ Hz for the following four classes ($G=4$): ``Idle,'' ``Snake,'' ``Updown,'' and ``Wave.'' In their respective order, the classes represent scenarios where the board on which the accelerometer is fixed remains stationary with occasional movement detected, glides across a desk in a way that resembles the movement of a snake, oscillates in a continuous up and down motion, and sways back and forth horizontally mimicking the gesture of waving to someone. The data are processed into $4,150$ samples ($3,527$ for training and $623$ for testing), each containing $K=15$ features obtained from the short-time Fourier transform of the acceleration time histories, as illustrated in Fig.~\ref{Fig6}a. The MNIST digit recognition dataset consists of $70,000$ samples ($60,000$ for training and $10,000$ for testing), each containing $K=784$ features obtained from $28 \times 28$ pixel grayscale images of hand-written digits, as shown in the same figure. This dataset has $S=10$ different classes corresponding to each integer digit from $0$ to $9$. The modulation frequencies in the first and second metasurface layers are $\omega_m = 7$ kHz and $\omega^{\prime}_m = 2$ kHz, respectively. The modulation depth is set to $\delta = \delta^{\prime} = 0.15$ in both layers. 

Consistent with Sec.~\ref{subsection_metasurface_parallel_classifications}, the gesture detection classification task utilizes the first neural network channel, where the input features are encoded at $\omega_1 = 150$ kHz, and the detection plane is partitioned into $G = 4$ segments. For each segment $g \in [1,G]$ corresponding to a gesture class, a displacement vector $\smash{\mathbf{u}_{6, g}^{[152 \; \mathrm{kHz}]}}$ is defined over a set of detection points. The intensity associated with each class is calculated as $\smash{|\mathbf{u}_{6, g}^{[152 \; \mathrm{kHz}]}|}^2 = {\mathbf{u}_{6, g}^{[152 \; \mathrm{kHz}]}}^{*} \cdot \mathbf{u}_{6, g}^{[152 \; \mathrm{kHz}]}$, where superscript $^{*}$ denotes the complex conjugate. The class corresponding to the segment with the highest intensity marks the predicted label. In parallel, the MNIST digit recognition classification task utilizes the second neural network channel. In this case, the input features are excited at $\omega_2 = 142$ kHz, and the corresponding detection plane is divided into $S = 10$ segments, each associated with a digit class. Similar to the first task, the intensities of the wave displacement vectors within each segment are evaluated, and the segment with the highest energy indicates the predicted class.

The goal of training is to identify $m$, $k_0$, $\phi_m$ for each and every waveguide in the first metasurface layer, as well as $m^{\prime}$, $k^{\prime}_0$, $\phi^{\prime}_m$ for each and every waveguide in the second metasurface layer. To train the models, a custom feed-forward neural network is implemented based on the scheme outlined in Sec.~\ref{subsection_metasurface_scattering} using \textsc{tensorflow keras}. An \textsc{adam} optimizer is adopted to backpropagate the error and minimize the loss function in $70$ epochs. After training, the values of $m$, $k_0$, $\phi_m$, $m^{\prime}$, $k^{\prime}_0$, and $\phi^{\prime}_m$ are obtained and employed to design all the waveguides across the neuromorphic system.

\subsection{Results and performance}
\label{subsection_metasurface_results}

Following the training and design process detailed earlier, the capability of the time-modulated neuromorphic metasurface to execute both tasks simultaneously is assessed in Fig.~\ref{Fig6}. The confusion matrix corresponding to the gesture detection classifier (results traced along the frequency path $150\; \mathrm{kHz} \rightarrow 150\; \mathrm{kHz} \rightarrow 152\; \mathrm{kHz}$) is depicted in Fig.~\ref{Fig6}b, whereas the confusion matrix corresponding to the MNIST digit recognition classifier (results traced along the frequency path $142\; \mathrm{kHz} \rightarrow 149\; \mathrm{kHz} \rightarrow 149\; \mathrm{kHz}$) is shown in Fig.~\ref{Fig6}c. Both matrices clearly capture the effectiveness of both computational channels and the high accuracy with which both classifiers operate subject to the test data from both datasets. Maximum accuracies achieved for the two tasks were found to be $93\%$ and $87\%$, respectively.

To demonstrate the operation of the trained system in the inference phase, two test samples are selected, one from the gesture detection dataset associated with the class ``Wave,'' and another from the MNIST dataset associated with the class ``$6$.'' Recalling that features of both classes were inputted to the system via mechanical excitations at $150$ and $142$ kHz, respectively, the contour plot at the rightmost side of Fig.~\ref{Fig6}a portrays the Multifrequency wave scattering pattern as a result of such excitations. Consequently, the shown plot displays the aggregate wavefield of all the resulting scattering channels, i.e., 18 frequency channels; nine for each input frequency component (in other words, the nine channels visualized in Fig.~\ref{Fig5}a, repeated for the input frequencies $\omega_1$ and $\omega_2$). To provide a closer look at the two computations running in parallel, frequency-domain snapshots are obtained from the aforementioned wavefield via fast Fourier Transform, which enables us to selectively observe the wave propagation pattern at the two frequency paths corresponding to the channels hosting the two neural networks. In Fig.~\ref{Fig6}b, the wavefield that traces the path $150\; \mathrm{kHz} \rightarrow 150\; \mathrm{kHz} \rightarrow 152\; \mathrm{kHz}$ is displayed, showing the neuromorphic metasurface's behavior as it tackles the gesture detection task. As expected, most of the energy at the detection plane focuses at the vertical segment corresponding to the correct label ``Wave.'' On the other hand, the wavefield tracing the path $142\; \mathrm{kHz} \rightarrow 149\; \mathrm{kHz} \rightarrow 149\; \mathrm{kHz}$ is shown in Fig.~\ref{Fig6}c, where the neuromorphic metasurface conducts an MNIST digit recognition. The figure again signals the physical system's computational success, and a focused energy beam can be observed at the location corresponding to ``$6$'' along the detection plane. The effectiveness of both outcomes is further confirmed by numerically evaluating the distribution of intensities over the detection planes for both cases, which are projected alongside the contour plots as discrete sidebars for reference.

While the metasurface system achieves high classification accuracy for the current tasks ($93\%$ and $87\%$ for the gesture detection and MNIST digit recognition tasks, respectively), the system's performance could be improved by enhancing the amount of energy focused at the target regions and reducing energy leakage to nontarget regions. In this work, we exclusively employ unit cells that provide relatively high transmission amplitudes. This choice is motivated by the desire to maximize energy output at the readout plane while avoiding the complications brought by wave reflections that can arise when using unit cells with low transmission amplitudes. However, as shown in Fig.~\ref{Fig4}e, the unit cells support a broad range of transmission amplitudes and phases. Leveraging this full range could offer greater design flexibility and enable metasurface configurations with potentially improved accuracy and overall performance. Additional strategies include imposing additional constraints on the neural network designing the metasurfaces, such as penalty terms that discourage side scattering, as well as exploring alternative optimizers, learning schemes, and regularization techniques. Finally, advanced neural architectures like Fourier neural operators \cite{li2020fourier}, which are tailored for solving partial differential equations and wave problems, represent another promising avenue for further performance gains.
\section{conclusions} 
\label{section_conclusions}

In summary, this work presents a successful realization of a dual classifier neuromorphic metasurface, addressing a major limitation of physical computing systems, namely the inability to concurrently carry out operations in parallel. The system shown here is capable of handling two independent intelligence tasks and has been demonstrated to do so with a minimal trade-off in effectivity and accuracy. The dual classifier working concept relies on the premise that by interjecting a temporal modulation in the physical profile of the metasurface layers (in here, the stiffness of the encompassing waveguides was chosen as a testbed), a set of additional scattering frequencies emerge that can be strategically utilized as distinct computational channels. Despite the fact that the dynamically changing stiffness results in the onset of a multitude of interlinked harmonics, we have demonstrated that a tactical choice of frequencies that house the computations pertaining to each task needs to be made in order to establish truly independent computational lanes, mitigating the risk of crosstalk and operation failure. This is done through adequate tuning of the modulation parameters (e.g., phase) based on a thorough understanding of the scattering harmonics generated by the different time-modulated layers. The framework shown here eliminates the need to physically replicate single-task metasurfaces to realize parallel computations, therefore significantly reducing footprint, cost, and fabrication challenges. 

Beyond dual-task functionality, it is important to emphasize that such a neuromorphic metasurface is inherently scalable, and that additional computational channels can be realized through appropriate tuning of the design parameters. Furthermore, the proposed framework does not rely on a specific unit-cell geometry. Instead, it adopts a lumped parameter approach, allowing the use of various unit-cell configurations, of which there are many in the metamaterial space, tailored to specific applications or criteria including operational frequencies, size, or material constraints. Finally, we note that while the primary focus of this paper is on wave propagation in elastic substrates, the fundamental principles established herein are broadly applicable owing to the analogous constitutive equations across different wave-based environments. As such, the insights shed here readily extend to acoustic and optical systems, effectively advancing the development of efficient multifunctional neuromorphic systems for physical intelligence.

\begin{acknowledgments}
This work was supported by the Mechanical Behavior of Materials program of the U.S. Army Research Office (ARO), under Grant No. W911NF-23-1-0078.
\end{acknowledgments}

\section*{Data Availability}
The data are available from the authors upon reasonable request.

\appendix* \section{Transfer matrix of a time-modulated unit cell}

Defined in Eq.~(\ref{tf}), the full structure of the transfer matrix of a time-modulated unit cell is given by
\begin{widetext}
\begin{equation}
    \mathbf{T} =
\begin{bmatrix}
        1 - \frac{m {\omega^-}^2}{2 k_0} & \frac{1}{k_0} - \frac{m {\omega^-}^2}{4 k^2_0} & \frac{m \delta \omega^2 e^{-\ii \phi_m}}{4 k_0} & \frac{m \delta e^{-\ii \phi_m}[{\omega^-}^2 + \omega^2]}{8 k^2_0} - \frac{\delta e^{-\ii \phi_m}}{2 k_0} & 0 & 0
        \\[0.2cm]
        
        -m {\omega^-}^2 & 1-\frac{m {\omega^-}^2}{2 k_0} & 0 & \frac{m \delta e^{-\ii \phi_m} {\omega^-}^2}{4 k_0} & 0 & 0
        \\[0.2cm]
        
        \frac{m \delta {\omega^-}^2 e^{\ii \phi_m}}{4 k_0} & \frac{m \delta [{\omega^-}^2 e^{\ii \phi_m} + \omega^2]}{8 k^2_0} - \frac{\delta}{2 k_0} & 1 - \frac{m \omega^2}{2 k_0} &\frac{1}{k_0} - \frac{m \omega^2}{4 k^2_0} & \frac{m \delta {\omega^+}^2 e^{-\ii \phi_m}}{4 k_0} & \frac{m \delta [{\omega^+}^2 e^{-\ii \phi_m} + \omega^2]}{8 k^2_0} - \frac{\delta}{2 k_0}
        \\[0.2cm]
        
        0 & \frac{\delta (e^{\ii \phi_m}-1)}{2} + \frac{m \delta \omega^2}{4 k_0}  & -m \omega^2 & 1 - \frac{m \omega^2}{2 k_0} & 0 & \frac{\delta (e^{-\ii \phi_m}-1)}{2} + \frac{m \delta \omega^2}{4 k_0}
        \\[0.2cm]
        
        0 & 0 & \frac{m \delta \omega^2 e^{\ii \phi_m}}{4 k_0} & \frac{m \delta e^{\ii \phi_m} [{\omega^+}^2 +  \omega^2]}{8 k^2_0} - \frac{\delta e^{\ii \phi_m}}{2 k_0} & 1 - \frac{m {\omega^+}^2}{2 k_0} & \frac{1}{k_0} - \frac{m {\omega^+}^2}{4 k^2_0}
        \\[0.2cm]
        
        0 & 0 & 0 & \frac{m \delta e^{\ii \phi_m} {\omega^+}^2}{4 k_0} & -m {\omega^+}^2 & 1-\frac{m{\omega^+}^2}{2 k_0}
\end{bmatrix}
\end{equation}
\end{widetext}

\nocite{*}

\bibliography{references}

\end{document}